\documentclass{article}

\usepackage{arxiv}

\usepackage[utf8]{inputenc} 
\usepackage[T1]{fontenc}    
\usepackage{hyperref}       
\usepackage{url}            
\usepackage{booktabs}       
\usepackage{amsfonts}       
\usepackage{nicefrac}       
\usepackage{microtype}      
\usepackage{lipsum}
\usepackage{graphicx}
\usepackage{threeparttable}
\usepackage[T1]{fontenc}

\title{(Unintended) Consequences of export restrictions on medical goods during the Covid-19 pandemic}
\date{}

\author{
 Marco Grassia\\
 Dip. Ingegneria Elettrica, Elettronica e Informatica\\
 Universit\`{a} degli Studi di Catania, Italy\\
 \texttt{marco.grassia@unict.it}
 \And
 Giuseppe Mangioni \\
 Dip. Ingegneria Elettrica, Elettronica e Informatica\\
 Universit\`{a} degli Studi di Catania, Italy\\
 \texttt{giuseppe.mangioni@unict.it}
 \And
 Stefano Schiavo \\
 Scuola di Studi Internazionali\\
 Universit\`{a}  di  Trento, Italy \\
 \texttt{stefano.schiavo@unitn.it}
 \And
 Silvio Traverso \\
 Scuola di Studi Internazionali \\
 Universit\`{a}  di  Trento, Italy \\
 \texttt{silvio.traverso@unitn.it}\\
}
 
\date{}

\renewcommand{\headeright}{}
\renewcommand{\undertitle}{}

\begin{document}

\maketitle

\begin{abstract}
In the first half of 2020, several countries have responded to the challenges posed by the Covid-19 pandemic by  restricting  their export of medical supplies.
Such measures are meant to increase the domestic availability of critical goods, and are commonly used in times of crisis. 
Yet, not much is known about their impact, especially on countries imposing them. 
Here we show that export bans are, by and large, counterproductive. Using a model of shock diffusion through the network of international trade, we simulate the impact of restrictions under different scenarios. We observe that while they would be beneficial to a country implementing them in isolation, their generalized use makes  most countries  worse off relative to a no-ban scenario. As a corollary, we estimate that prices increase in many countries imposing the restrictions. We also find that the cost of restraining from export bans  is small, even when others continue to implement them. 
Finally, we document a change in countries' position within the international trade network, suggesting that export bans have geopolitical implications.
\end{abstract}


\section*{Introduction}

During the Covid-19 pandemic, several countries have resorted to non-cooperative trade policies in the form of restrictions to the export of essential  medical supplies.
These kind of measures are meant to insulate the domestic market from a shock (being it internal or external), limiting  expected shortfalls in the availability of goods that may result from either increased demand or reduced supply.
Even if the track record of export restrictions is not immaculate, the most recent example being the 2007--2008 spike in the price of rice, this policy tool remains very popular. 
So much so, that  in the first  months of 2020, more than 50 governments have imposed some  curbs on the exports of medical supplies, ranging from licensing requirements to outright export bans \cite{bown:2020, gta2020}.

This is a typical instance in which economic theory, which predicates the efficiency-enhancing virtues of free trade, clashes both with  common sense  and  with the political need to  \emph{do something} \cite{egan:2014, evenett:2020b}. 
Yet, are these measures effective? 
This is not just an abstract academic question. Rather, it has profound and general implications both in terms of how countries react to a pandemic and, more generally, how they can adequately respond to global crises. 

Existing studies highlight the distributional impact of beggar-thy-neighbor non-cooperative trade policies, with developing and low-income countries posed to suffer the most from export restrictions \cite{evenett:2020,rutaC:2020} and policy prescriptions recommending import-dependent countries to diversify their supplies in order to cope with exogenous disruptions.   
Others  describe the mechanisms through which a \emph{domino effect} can unravel, whereby export curbs by a single country will exacerbate shortages and thus  increase the incentive for other governments to follow suit  \cite{giordaniC:2016}. 

In this study we move forward and ask whether export restrictions provide countries imposing them with any meaningful benefit. 
We do so by using a model of shock diffusion through a network where countries are nodes and export flows represent links among them. 
A network approach grants us the  ability to look beyond the impact of restrictions on a country's direct trade partners, and to study the indirect effect on third countries, as well as at feedback loops on initiating countries\cite{barigozzi2011identifying, alves2019nested}. Both effects crucially depend on  the complex network of bilateral relations which characterizes international trade. 
 
We show that, by and large, export bans are not effective. 
The vast majority of countries that impose restrictions end up with a demand deficit even after accounting for the local availability of goods that were previously sold abroad. 
Similarly, most countries that adopt export curbs experience price increases relative to a \emph{business-as-usual} (BAU) scenario. 
What is more, a counterfactual analysis shows that even if individual countries restrained from imposing restrictions while others continued to do so, very few of them would experience a sharp deterioration of their position.
This suggests that there is very little economic rationale for adopting these kind of measures and countries should avoid getting caught into a \emph{restriction frenzy}.

Our findings are all the more remarkable in that they abstract from  global value chains (and the ensuing dependence of domestic production on foreign inputs), or from the surge in demand for medical goods triggered by the pandemic. While both these factors would exacerbate the negative effects of import restrictions, our setting is useful to isolate and understand the effects of the simple \emph{zero-sum logic} behind  export bans.  Similarly, we do not consider any increase in domestic production capacity, since this is unlikely to be quantitatively relevant in the short run. 

The results have implications that are not limited to economics, but touch upon  the position and role of countries within the international arena.  
By imposing export bans,  countries somehow cut themselves off global trade (even if for a limited period and a narrow range of goods): this  may have geopolitical effects as it leaves space to be filled by other countries \cite{javorcik:2020, weinhardt_tenbrink:2020, wu:2016,doshi:2020}. 
We investigate this phenomenon by looking at measures of network centrality, and  document that this is indeed what has happened in the case of the United States (whose centrality has declined) and China (that has improved its standing in the trade network). 

While our work contributes to the growing literature that investigates, almost in real time, the economic impact of  Covid-19 \cite{baldwindimauro:2020,cepr_book2020}, its approach can find applications above and beyond the current pandemic.   
Indeed, the simulations  pave the way for a better understanding of the  interactions that characterize the international trade system, that is necessary to assess the impact of national and international policies aimed at providing an effective response to the next global crisis.

\section*{Results}

The analysis combines data on bilateral trade flows with information on export restrictions on medical goods to investigate their effects on the countries imposing them, their direct partners, and third countries.

\subsection*{Main findings}
As reported in Table \ref{tab:overall},  the amount of trade concerned by restrictions varies between less than 1\%  of global exports in the case of consumable medical goods (with only 4 countries imposing restrictions) to 21.5\% when it comes to protective garments and disinfectants (whose exports have been stopped by 29 and 19 countries).

While simulations have been run on all categories of medical goods that are relevant in the fight against the new coronavirus, in what follows we will mainly discuss results for  ``protective garments'', as this is the category mostly hit by export restrictions, both in terms of countries imposing curbs  and  of the share of exports covered. 

In general, two broad patterns emerge from our baseline simulations.
First of all,  export bans on medical equipment  imposed by just few governments affect several  countries, even those that are not directly sourcing from the initiators. 
The impact is often severe, with a relatively large  number of countries (and share of world population) no longer able to import  medical supplies. 
In this regard, the case of \emph{test kits} is emblematic: while only two countries impose bans, the United Kingdom and Belgium, their combined market share is about  10\%; as a consequence, 79 countries (29.4\% of world population) see their imports falling by 75\% or more (with 62 countries becoming unable to import from their usual sources), and  37 others experience a reduction of at least 25\% (see Table \ref{tab:overall}).
Complex diffusion dynamics due to the network structure imply that even though the number of heavily affected  countries tends to increase with the share of total trade restricted, also limited export bans, such as those on \emph{consumables}, \emph{soap}, and \emph{other medical devices}, can produce sizeable effects on several  countries.

Second, not all the countries imposing bans benefit from them. 
In the case of \emph{protective garments}, for instance, 18 of the the 28 countries that curb exports  are worse-off relative to the BAU scenario (see Table \ref{tab:cat2}). 
This is linked to the observation that  most of the countries that adopt restrictive measures export less than they import: this is the case for 22 out of the 28 countries under  consideration, which display an average  export-to-import ratio equal to 0.32. 
Moreover, this  feature is common to all the goods we analyze. 
The US represents an interesting case in point, as the value of its exports of \emph{protective garments} is only 10\% of the value of its imports.
From this, it is clear that initiating countries may be vulnerable to export restrictions imposed by other governments that decide to retaliate, or that are simply dragged into limiting exports by the fear to appear weak in the eye of the public or by sheer panic \cite{giordaniC:2016}.


In fact, the simulations indicate that --apart from the case of test kits, where the two countries restricting exports manage to increase the domestic availability of the product relative to the BAU scenario-- many of the countries imposing a ban end up with a net demand deficit. In other words, even considering that exports are restricted and thus the goods previously shipped abroad are now  available for domestic consumption, these countries are no longer able to import all the goods they need (absent any increase in domestic demand). In particular, the share of countries that impose a ban but turn out to be worse-off ranges from 20\% (\emph{soap}) to more than 60\% (\emph{protective garments} and \emph{consumables}). 

\begin{figure}[htb]
\includegraphics[width=\linewidth]{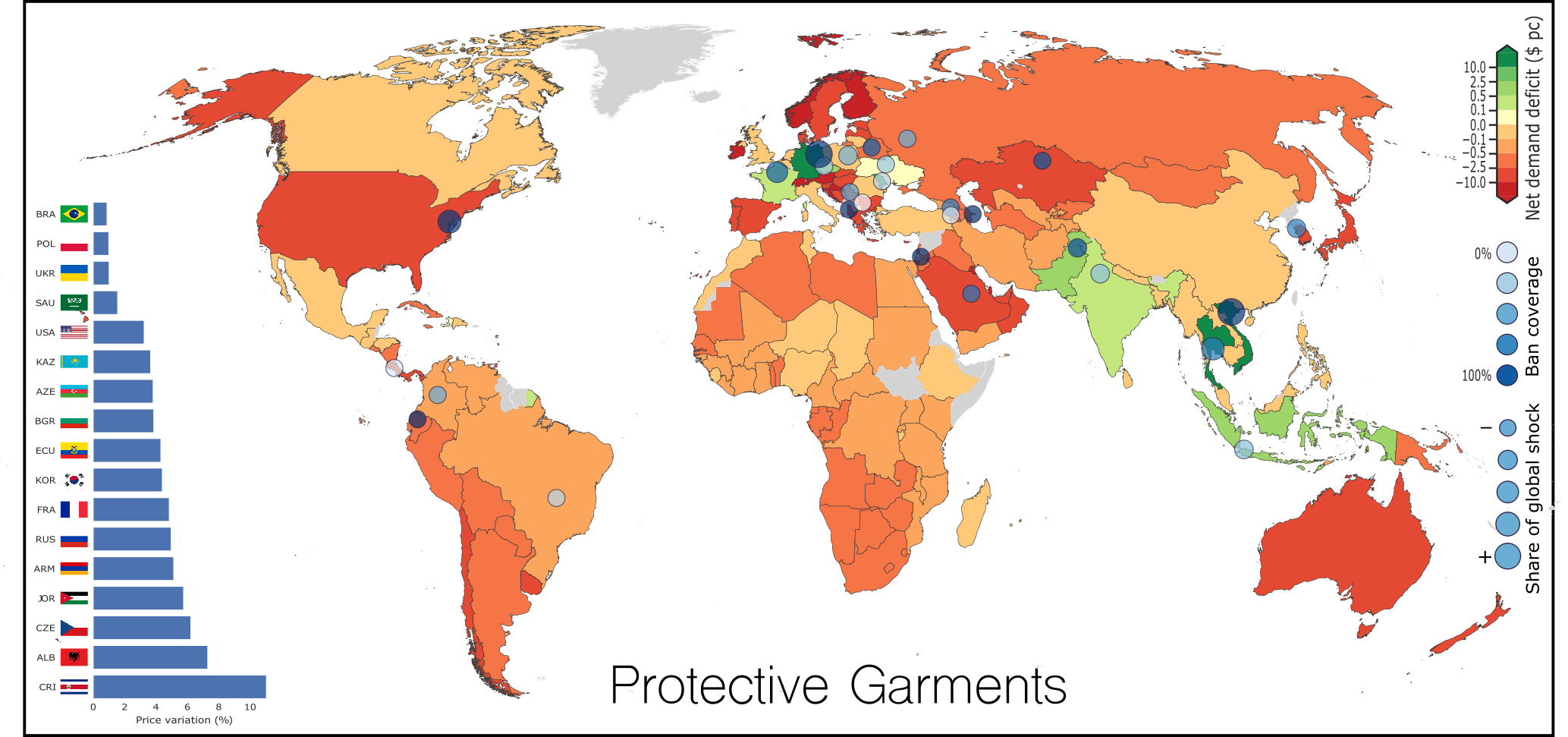}
\caption{Impact of export bans on protective garments.  For each country, the color indicates the value of the net demand deficit in USD per capita. The size and color of  dots on the capital city of countries imposing restrictions represent, respectively, the contribution to the total shock across the world and the percentage of the country's exports that are banned (if any). The bar-plot on the left shows the price increase in the countries that impose a ban and experience a net demand deficit. Price changes are computed multiplying the expected reduction in the available quantity of protective garments by the relevant import price elasticity (estimated by Fontagn\'e and coauthors\cite{fontagneC:2019}).}
\label{fig:map}
\end{figure}

Figure \ref{fig:map} provides a graphical representation of the effect of export bans on world countries in the case of protective garments, showing that many of those imposing restrictions end up with significant shortages of goods.
Two large countries (severely affected by the pandemic) for which the  ban on protective garments is counterproductive are the US and Russia. 
Both  have curbed exports despite shipments  abroad represent just about 10\% of  imports, and the simulations predict they will experience a  net demand deficit amounting to  12\% of initial imports for the US and  28\% for Russia. 
On the other hand, France, Germany, the Czech Republic and Ukraine (net importers) plus all net exporters  feature a surplus of protective garments.
The driving force behind the difference in the outcome appears to be the export-to-import ratio, which is more than twice as large (0.59 vs. 0.26) for the net importers benefiting from the ban (let aside the net exporters).

\subsection*{Counterfactual simulations}

To  assess the impact of export bans on individual countries, we run a series of counterfactual simulations in which (i) we simulate the effect of an export ban implemented by a single country; or (ii) we exclude a single country from the list of those imposing the restrictions. 

In the first case, all countries benefit from curbing exports. Because trade partners cannot reciprocate, the worst case scenario is one where the entire shock is absorbed by the initiating country, which would then end up in the same position as in the BAU case.
In fact, we observe that all the 28 countries restricting exports of protective garments would either increase their net surplus or move from a net deficit to a net surplus (see column 2 of Table \ref{tab:itself}). The US, for instance, would jump from a shortfall equal to  11\% of its imports to an ``excess supply" of  6\% of imports.
Overall, when comparing  columns 1 and 2 of  Table \ref{tab:itself}, it is clear that  a policy measure that might work when implemented in isolation, is very often counterproductive  when its adoption is  widespread.  
In our simulations for protective garments this happens to  18 out of 28 countries.

If that is the case, does it mean that  countries are cajoled into adopting restrictive trade policies by others' behavior? In other words, is this just a  ``bad equilibrium" stemming from strategic interaction,  a situation that resembles the well-known ``prisoners' dilemma" setup? 

Column 3 of Table \ref{tab:itself} shows that for almost all the countries, playing a ``cooperative strategy" while others ``defect" (to stick to game-theoretical jargon) would not lead to a negative outcome. 
While some countries are actually penalized, these are the net exporters, which  move from experiencing a net surplus in the baseline simulation  to zero, thus suffering no  deterioration with respect to the initial BAU scenario.  
The only one that would experience a negative effect is Russia, but  the additional negative effect is very limited, as it demand deficit increases from 28.3 to 28.5\% of initial imports.
Overall then, the costs of a cooperative strategy are trivial.

\subsection*{Additional Results}
Two additional results are worth noting, as they provide us with alternative metrics to assess the impact of export restrictions on the countries imposing them: changes in prices and in network centrality.

\subsubsection*{Price effect}

Using the change in import quantities (relative to the 2018 data) that are predicted by the simulations for each country, and import elasticities (that is, the sensitivity of prices to changes in import quantities) available for each product \cite{fontagneC:2019} we can assess the impact of export restrictions on prices.

As detailed in Table \ref{tab:prices},  the average increase in import prices ranges between 0.4\% for soap to 8.9\% in the case of other medical devices, but there is a lot of cross-country heterogeneity.
The  countries that manage to increase the domestic availability of  goods clearly  enjoy (everything else equal, that is, absent any increase in demand) a reduction in prices. 
If we focus on protective garments, the average price reduction  (-35.3\%) masks a great deal of heterogeneity even within this small group (10 countries). 
In fact, large exporters imposing export restrictions such as Vietnam, Thailand, Indonesia and Pakistan,  experience very large fall in prices (up to 100\%, that is, driving foreign goods out of the market), while for the other countries the reduction hovers around -10\%.   

This contrasts with an average increase of 7.5\% for all the countries that do not adopt protectionist measures (the impact ranges between 0 and 33\% for them) and an increase of up to 11\% for the 17 countries that, despite enacting export bans, end up with a net demand deficit. 
These numbers, in particular those relative to the effect on import prices in countries that impose trade restrictions,  represent yet another measure of the performance of export bans and show that uncooperative trade policies may well have unintended negative effects even on the countries that implement them, above and beyond the adverse effects they impose on other economies.

\subsubsection*{Centrality}
A last interesting element emerges from the comparison of the pre- and post-shock network structure, that is, the set of bilateral flows emerging from the simulations incorporating export bans relative to the actual trade links observed in the 2018 data. 

Because the analysis assumes the elimination of a set of links between the initiating countries and (some of) their partners, global connectivity will decline.
We are interested in  whether countries that impose export restrictions undergo any significant change in their position within the network compared to the rest of world.
To investigate this hypothesis, we compare the change in the centrality (hub) scores of countries imposing export bans before and after the shock with the variation registered for other countries.

The 28 countries that impose an export ban on protective garments witness a reduction in their hub score that is significantly larger that the average fall in centrality, and larger that the one observed for countries that have not enacted restrictive measures. 
This is confirmed  both by a two-sample $t$-test that compares the means for the two groups of countries (equality of means is rejected with a $p-value = 10^{-4}$), and a regression  of the difference in the hub-scores (computed on the simulated vs. the original network) on an indicator variable that takes value one for countries adopting the ban (the indicator has a negative coefficient $-0.01$ that is significantly different from zero with a $p-value = 6.6 \cdot 10^{-5}$). 

Looking at specific countries, we see that China, which tops the ranking in terms of centrality,  is one of the very few countries that further increase their hub score (it is actually the country whose score increases the most), while Vietnam and Thailand, two large exporters of protective garments that have imposed an export ban, experience a fall in centrality. Similarly, the US, which has imposed a ban even if its exports is limited, sees its position within the network sliding further.

In this respect, speculation that the different attitudes of countries in terms of international collaboration during the Covid-19 pandemic may have geo-strategic resonance \cite{doshi:2020} by altering their positions within the global trade network, seems to find some comfort in the data.

\section*{Discussion}

We have shown  that export restrictions in times of crises are by and large counterproductive. 
Our simulations suggest that  most countries imposing a ban face lower availability of goods and higher prices. Moreover, they  would not lose much in case they avoided restrictions, even if other countries were still implementing them. 

Three main reasons stand behind these results. 
First and foremost, because the shortage of medical goods is global, restricting exports does not  address the roots of the problem as it does not increase overall supply.
Second, trading partners can retaliate against  unilateral decisions, triggering a domino effect  that may well backfire. 
Third, because international trade is organized as a complex network of bilateral connections, it is  difficult to predict the consequences associated with removing even a few trade links from the system. 

The results presented in this work shed  light on the (often unintended) consequences of non-cooperative trade policy. As such, their implications go beyond the measures adopted during the Covid-19  pandemic and offer an evidence-based contribution to the  debate on the best practices to adopt during global crises.

\section*{Methods}

\subsection*{Data}
The analysis combines data on international bilateral trade flows as reported in the CEPII-BACI dataset \cite{baci:2010} with information on export restrictions collected by Global Trade Alert \cite{gta2020}. 
Bilateral trade data are reported in 1,000 dollars and metric tons. 
Data on GDP and population are taken from the World Bank \cite{wb:2020}.
Finally, import price elasticity necessary to compute price changes are taken from \cite{fontagneC:2019}.

We focus on export bans implemented between January 1st and April 30th, 2020 that concern a list of 32 Covid-relevant products organized in six categories.
The product groups  are: test kits, protective garments, disinfectants, other medical devices, consumables, and soap. A seventh category, thermometers, features no export restrictions and is therefore excluded from the analysis. 
For a detailed list of relevant products comprised in each category, see \cite{gta2020}. We do not consider other forms of export restrictions such as licensing requirements.
For each initiating country, that is countries imposing export bans, we identify the specific category of goods concerned and the trade partners affected by the restriction.

\subsection*{International trade networks of medical goods}
International trade data for year 2018 (the latest available) are used to build a BAU scenario that represents the pre-shock reference point. Using countries with a population of at least one million as nodes and bilateral trade flows as links, we build a weighted and directed network for each of the six product categories and let a shock diffuse through each of them, originating from  the countries imposing an export ban.

In formal terms, the network of each product category $p$ is represented by a weighted directed graph $G^p = (V^p, E^p, W^p)$, where $V^p=\{c_i: i \in \{1, ..., N\} \}$ is a set of nodes ($N=148$), $E^p = \{(c_i, c_j): i,j \in \{1, ...,N\} \}$ is a set of directed edges between pairs of nodes, and $W^p = \{W^p_{c_ic_j}: i,j \in \{1, ...,N\} \}$ is the set of the weights associated with the edges (i.e., the monetary value of the export of product $p$ from country $c_i$ to country $c_j$). 

\subsection*{Diffusion model}
The adoption of export restrictions by certain countries reduces the availability of goods for their trade partners and represents the initial shock, which then propagates  through the  trade network. 
In particular, countries facing a shortage of imported goods will  (try to) compensate  by reducing their own exports \cite{burck2019}. This implies that domestic demand has  higher priority relative to foreign demand. When exports cannot be reduced further, the country registers a \emph{demand deficit}, that is situation whereby  the amount of medical goods available in the country is lower than in the BAU scenario. 

The model assumes that no new bilateral trade relationship can be established in the short run and that the reduction in exports will affect partners with a magnitude that is inversely proportional to their economic size, measured in terms of GDP . 
This second assumption represents a major departure from previous studies \cite{burck2019} and is meant to capture the differences in countries' purchasing power\cite{distefanoetal:2018}. Indeed, poor countries are likely to be disproportionately affected by global shortages \cite{evenett:2020}.

To formally introduce our shock diffusion model, we first define the equilibrium domestic demand (i.e., before the shock) of the generic county $c_i$ for product $p$ as 
\begin{equation}
dem^p_{c_i}(t) = prod^p_{c_i}(t) + imp^p_{c_i}(t) - exp^p_{c_i}(t)
\end{equation}
in which $prod^p_{c_i}$, $exp^p_{c_i}$, and $imp^p_{c_i}$ indicate domestic production, export and import of product $p$ respectively. Note that in network terms (and abstracting from time $t$) 
\begin{equation}
 exp^p_{c_i} = \sum_{j = 1}^N W^p_{c_ic_j} 
\end{equation}
\begin{equation}
imp^p_{c_i} = \sum_{j = 1}^N W^p_{c_jc_i}.
\end{equation}
Although we do not observe domestic production  $prod^p_{c_i}$, we assume it to be constant in the short run.

A generic country $c_s$ that imports from a partner $c_r$ imposing an export ban  will face a shock equal to the amount of the bilateral import flow  and, everything else equal, a demand deficit of the same magnitude: 
\begin{equation}
dd^p_{c_s} = \sum_{r = 1}^N  b^p_{rs} \cdot W^p_{c_rc_s}  
\end{equation}
where $b^p_{rs}$ is an indicator variable taking value 1 if country $r$ imposes a ban on exports of product $p$ to country $s$, and zero otherwise.
According to our model, $c_s$ will then try to offset this demand deficit by reducing its own export by the same amount. Thus, in the next step, the new level of export of $c_s$ will be

\begin{equation}
exp^p_{c_s}(t) = max\{exp^p_{c_s}(t-1) - dd^p_{c_s}(t), \; 0\}.
\end{equation}

This, in turn, induces a cascading effect on those countries that  import from $c_s$.
In fact, after the initial step, the shock propagates in the network producing a demand deficit in a generic country $c_i$ at time step $t$ given by 
\begin{equation}
dd^p_{c_i}(t) = dem^p_{c_i}(t) - prod^p_{c_i}(t) - imp^p_{c_i}(t) + exp^p_{c_i}(t).
\end{equation}
It is easy to see that if imports and exports do not change, the demand deficit equals zero $dd^p_{c_i}(t) = 0$.

While at the beginning of the simulation the reduction of exports mimics the actual policy choices of countries imposing the ban, in the following steps shocks spread in a way that is inversely proportional to the GDP  of out-neighs. 
Formally, if $dd^p_{c_i}(t) > 0$, the reduction in  exports will be distributed among the countries that import from $c_i$ as

\begin{equation}
W^p_{c_ic_j}(t+1) = max \{ W^p_{c_ic_j}(t) - dd^p_{c_i}(t)\left[ \frac{ \sum_{h\neq j}  GDP_{c_h}}{\sum_h GDP_{c_h}}*\frac{1}{odeg_{c_i} -1} \right], 0 \}  
\end{equation}
where $GDP_{c_h}$ is the GDP  of the generic out-neighbor $c_h$ and $odeg_{c_i}$ is the out-degree of country $c_i$ (i.e., the number of outward edges departing from $c_j$).
The diffusion process stops  when no country facing  a positive demand deficit can further reduce its exports. 

In the case a country has imposed an export ban, its final demand deficit will be (partly or fully) compensated by the availability of goods that were previously shipped abroad:
\begin{equation}
    net\_dd^p_{c_r}(t) = dd^p_{c_r}(t) - \sum_{j=1}^N b^p_{rj} \cdot W^p_{c_rc_j}. 
\end{equation}


\subsection*{Prices}
In the case of net exporters that impose a ban to their sales abroad, we cap the reduction in import prices to -100\%, a value implying  that the increased domestic availability of goods that were previously exported would (everything else equal) drive imports basically out of the market. This happens to two countries (Thailand and Vietnam) for protective garments and to Costa Rica for other medical devices.






\begin{table}[htb]
  \centering
  \footnotesize
  \caption{Impact of export bans on net imports relative to a \emph{business-as-usual} scenario} \label{tab:overall}
  \begin{threeparttable}
    \begin{tabular}{lcccccccc}
\toprule
    &       & \multicolumn{7}{c}{Impact on net imports} \\
\cmidrule{3-9}         
    &  & reduction & reduction & reduction & no    & increase & increase & increase \\
Product category & coverage &  $>75\%$ & $25-75\%$ & $<25\%$ & effect & $<25\%$ & $25-75\%$ & $>75\%$ \\
    \midrule
Test kits & 2     & 79    & 37    & 7     & 23    & 0     & 1     & 1 \\
          & [9.9\%] & (29.4\%) & (33.7\%) & (4.0\%) & (31.8\%) & (0.0\%) & (0.9\%) & (0.2\%) \\
Protective garments & 29    & 35    & 53    & 20    & 32    & 2     & 0     & 6 \\
          & [21.5\%] & (6.6\%) & (15.8\%) & (15.7\%) & (33.9\%) & (1.3\%) & (0.0\%) & (26.8\%) \\
Disinfectants & 19    & 80    & 34    & 8     & 19    & 2     & 2     & 3 \\
          & [21.5\%] & (25.9\%) & (17.1\%) & (21.0\%) & (12.9\%) & (3.0\%) & (1.0\%) & (19.2\%) \\
Other medical devices & 8     & 50    & 37    & 11    & 46    & 1     & 0     & 3 \\
          & [3.2\%] & (12.2\%) & (13.6\%) & (10.2\%) & (42.5\%) & (18.1\%) & (0.0\%) & (3.4\%) \\
Consumables & 4     & 2     & 41    & 45    & 59    & 0     & 1     & 0 \\
          & [0.8\%] & (0.2\%) & (6.2\%) & (22.2\%) & (53.3\%) & (0.0\%) & (18.1\%) & (0.0\%) \\
Soap        & 5     & 2     & 7     & 51    & 84    & 2     & 1     & 1 \\
          & [1.2\%] & (0.4\%) & (1.8\%) & (10.5\%) & (67.0\%) & (1.4\%) & (18.1\%) & (0.7\%) \\
\bottomrule
    \end{tabular}%
    \begin{tablenotes}[para,flushleft] 		
Notes: the column \emph{coverage} indicates the number of countries which have imposed a ban and, in squared brackets, the share of total trade affected by the restrictions; the columns showing the \emph{Impact on net imports} report the number of countries and, in brackets, the share of world population affected; \emph{no effect} includes all the cases in which the absolute value of the variation is below 2\%. 							   \end{tablenotes} 
  \end{threeparttable}
\end{table}

\begin{table}[htb]
 \centering
 \footnotesize
 \caption{Impact of export bans on Protective Garments on countries imposing them}
  \label{tab:cat2}
\begin{threeparttable}
\begin{tabular}{lrrr}
\toprule
country code & export/import  & netDD (share of imports) & netDD (per capita)  \\
\midrule
ALB   & 0.512 & 0.44  & 2.545 \\
ARM   & 0.439 & 0.55  & 1.325 \\
AZE   & 0.005 & 0.758 & 1.414 \\
BGR   & 0.479 & 0.316 & 2.542 \\
BLR   & 0.374 & 0.446 & 2.116 \\
BRA   & 0.035 & 0.128 & 0.247 \\
COL   & 0.157 & 0.172 & 0.44 \\
CRI   & 0.051 & 0.703 & 6.577 \\
CZE   & 0.682 & -0.043 & -1.713 \\
DEU   & 0.673 & -0.234 & -10.793 \\
ECU   & 0.047 & 0.481 & 1.173 \\
FRA   & 0.448 & -0.006 & -0.204 \\
GEO   & 0.528 & 0.028 & 0.177 \\
IDN   & 3.603 & -1.962 & -1.246 \\
IND   & 2.091 & -1.146 & -0.246 \\
JOR   & 0.107 & 0.403 & 1.149 \\
KAZ   & 0.028 & 0.314 & 2.587 \\
KOR   & 0.348 & 0.131 & 2.526 \\
MDA   & 1.774 & -0.813 & -3.699 \\
PAK   & 4.107 & -3.723 & -1.353 \\
POL   & 0.56  & 0.001 & 0.021 \\
RUS   & 0.095 & 0.283 & 1.401 \\
SAU   & 0.057 & 0.373 & 3.29 \\
SRB   & 0.735 & 0.102 & 0.561 \\
THA   & 8.258 & -6.393 & -18.45 \\
UKR   & 0.550  & -0.018 & -0.041 \\
USA   & 0.109 & 0.116 & 4.756 \\
VNM   & 8.757 & -8.534 & -26.317 \\
 \bottomrule
 \end{tabular}%
    \begin{tablenotes}[para,flushleft] 		
\emph{Notes}. NetDD stands for the final demand deficit net of previously exported goods  that are available domestically due to the export bans. 
A negative figure implies a surplus, that is a situation where the domestic availability of goods is larger than in the BAU scenario. Per capita values in current USD. 
   \end{tablenotes} 
  \end{threeparttable} 
\end{table}

\begin{table}[hbt]
\caption{Impact of a country's behavior on its own demand deficit.}
 \label{tab:itself}
 \begin{center}
  \begin{threeparttable}
\footnotesize
\begin{tabular*}{.6\textwidth}{@{\extracolsep{\fill}}lrrr}
\toprule
country   & baseline & isolated ban & no ban \\
        & (1) & (2) & (3)    \\
\midrule
ALB   &   44.0\% &  -48.0\% & 44.0\% \\
AZE   &   75.8\% &   -0.3\% & 75.8\% \\
ARM   &   55.0\% &  -15.4\% & 55.0\% \\
BRA   &   12.8\% &   -1.6\% & 12.8\% \\
BGR   &   31.6\% &  -10.6\% & 31.6\% \\
BLR   &   44.6\% &  -31.4\% & 44.6\% \\
COL   &   17.2\% &   -9.6\% & 17.2\% \\
CRI   &   70.3\% &   -1.6\% & 70.3\% \\
CZE   &   -4.3\% &  -33.3\% &  0.0\% \\
ECU   &   48.1\% &   -4.2\% & 48.1\% \\
FRA   &   -0.6\% &  -37.9\% &  0.0\% \\
GEO   &    2.8\% &  -40.1\% &  2.8\% \\
DEU   &  -23.4\% &  -51.2\% &  0.0\% \\
IDN   & -196.2\% & -196.2\% &  0.0\% \\
KAZ   &   31.4\% &   -2.4\% & 31.4\% \\
JOR   &   40.4\% &  -10.0\% & 40.4\% \\
KOR   &   13.1\% &  -25.8\% & 13.2\% \\
MDA   &  -81.3\% &  -81.3\% &  0.0\% \\
PAK   & -372.3\% & -372.3\% &  0.0\% \\
POL   &    0.1\% &  -33.6\% &  0.1\% \\
RUS   &   28.3\% &   -5.9\% & 28.5\% \\
SAU   &   37.3\% &   -4.6\% & 37.3\% \\
SRB   &   10.2\% &  -47.9\% & 10.2\% \\
IND   & -114.6\% & -114.6\% &  0.0\% \\
VNM   & -853.4\% & -856.6\% &  0.0\% \\
THA   & -639.3\% & -639.3\% &  0.0\% \\
UKR   &   -1.8\% &  -28.8\% &  0.0\% \\
USA   &   11.6\% &   -5.7\% & 11.6\% \\
\bottomrule
\end{tabular*}%
\begin{tablenotes}[para,flushleft]
\emph{Notes}. The Table illustrates the net demand deficit for all countries implementing an export ban on protective garments. Values are in percentage of imports; negative numbers represent surpluses. 
Column (1) displays the results from the baseline simulation that considers export bans by all 28 countries. 
Column (2) presents simulations assuming that the export ban is implemented  by each   country in isolation. Results in column (3)  are based on the assumption that each country restrains from imposing a ban while the others continue to implement restrictive measures. 
\end{tablenotes}
  \end{threeparttable}
 \end{center}
\end{table}

\begin{table}[htb]
\caption{Price effects}
 \label{tab:prices}
 \begin{center}
  \begin{threeparttable}
\footnotesize
\begin{tabular}{lrrrrrrr}
\toprule
                    &       &          &        & \multicolumn{2}{c}{mean} \\
\cmidrule{5-6}                    
category            &  mean &  min$^{\dag}$ & max    & ban = 0 & ban = 1 \\
\midrule
test kits           & 7.2\% &  -15.2\% & 19.6\% &  7.5\%  & -14.6\% \\
protective garments & 4.0\% & -100.0\% & 33.1\% &  7.5\%  & -11.2\% \\
disinfectants       & 3.6\% &  -43.0\% & 30.4\% &  4.7\%  &  -3.8\% \\
other devices       & 8.9\% & -100.0\% & 25.9\% & 10.6\%  & -20.0\% \\
consumables         & 2.0\% &  -21.3\% & 15.2\% &  2.1\%  &  -3.7\% \\
soap                & 0.4\% &   -8.7\% & 11.3\% &  0.6\%  &  -3.1\% \\
\bottomrule
\end{tabular}%
 \begin{tablenotes}[para,flushleft]
 \emph{Notes}. [\dag] negative price variations are capped at -100\% in the case of large exporters imposing a ban. The amount of goods previously exported that are available for domestic consumption far exceeds imports, basically driving foreign goods out of the market. The constraint is biding for two countries in the protective garments category and for one country in other medical devices.  
 \end{tablenotes}
  \end{threeparttable}
 \end{center}
\end{table}

\end{document}